# Enhancement of photo-responsivity of small gap, multiple tunnelling superconducting tunnel junctions due to quasiparticle multiplication

G. Brammertz[1], A.G. Kozorezov[2], R. den Hartog[1], P. Verhoeve[1], A. Peacock[1], J.K. Wigmore[2], D. Martin[1], and R. Venn[3]

[1] Science Payload and Advanced Concepts Office, ESA/ESTEC, Postbus 299, 2200AG Noordwijk, The Netherlands.
[2] Department of Physics, Lancaster University, Lancaster, United Kingdom.
[3] Cambridge MicroFab Ltd., Trollheim Cranes Lane, Kingston, Cambridge CB3 7NJ, UK.

**Abstract.** We recently predicted the formation of a highly non-equilibrium quasiparticle (qp) distribution in low $T_C$ multiple tunnelling superconducting tunnel junctions (STJs) [1]. The situation arises through qp energy gain in cycles of successive forward and back tunnelling events in the absence of relaxation via sub-gap phonon emission. The qps can acquire sufficient energy to emit phonons, which break more Cooper pairs and release additional qps. In this paper we report theoretical and experimental studies of the effect of this process on photon detection by such an STJ. We derived a set of energy-dependent balance equations [2], which describe the kinetics of the qps and phonons, including the qp multiplication process described above. Solution of the balance equations gives the non-equilibrium distribution of the qps as a function of time and energy, and hence the responsivity of the STJ as a function of bias voltage. We compared the theoretical results with experiments on high quality, multiple-tunnelling Al STJs cooled to 35mK in an adiabatic demagnetisation refrigerator, and illuminated with monochromatic photons with wavelengths between 250 and 1000 nm. It was found that in the larger junctions with the longest qp loss time, both responsivity and signal decay time increased rapidly with bias voltage. Excellent agreement was obtained between the observed effects and theoretical modelling.

## 1. Introduction

We have recently developed high quality, low $T_C$ Al superconducting tunnel junctions (STJs) for application as optical to x-ray photon detectors in astronomy [3]. The motivation for this development is the increased theoretical resolving power as compared to higher gap Ta based junctions. The quasiparticle relaxation rate is approximately proportional to the cube of the energy gap $\Delta_g$ of the superconductor [4]. As a consequence, energy down-conversion of quasiparticles (qps) in these lower-$T_C$ superconductors becomes much slower and the bias energy gained by the qps due to successive tunnel and back-tunnel events leads to a very broad energy distribution of the non-equilibrium qp population created by the photon absorption process. We have therefore developed a model of the photon absorption process in STJs, which includes the full energy-dependence of all the qp processes occurring in these junctions. This model allows for the calculation of the time- and energy-dependent qp distribution in the electrodes of the detector.

## 2. Kinetics of the quasiparticle energy distribution in tunnel junctions used as photon detectors and quasiparticle multiplication rate

The developed model is based on a numerical approach, which divides the qp energy domain between $\Delta_g$ and $4\Delta_g$ into $n_{en}$ intervals of width $\delta\varepsilon$. Because of calculation time limitations, $n_{en}$ is generally chosen to be close to ~30. In the case of Al STJs, which are treated in this paper, $\Delta_g$ is equal to 180 µeV and $n_{en}$=33. As a consequence, the width of the basic energy interval $\delta\varepsilon$ is approximately equal to 16 µeV. Figure 1 shows a semiconductor representation of all the qp processes included in our model.

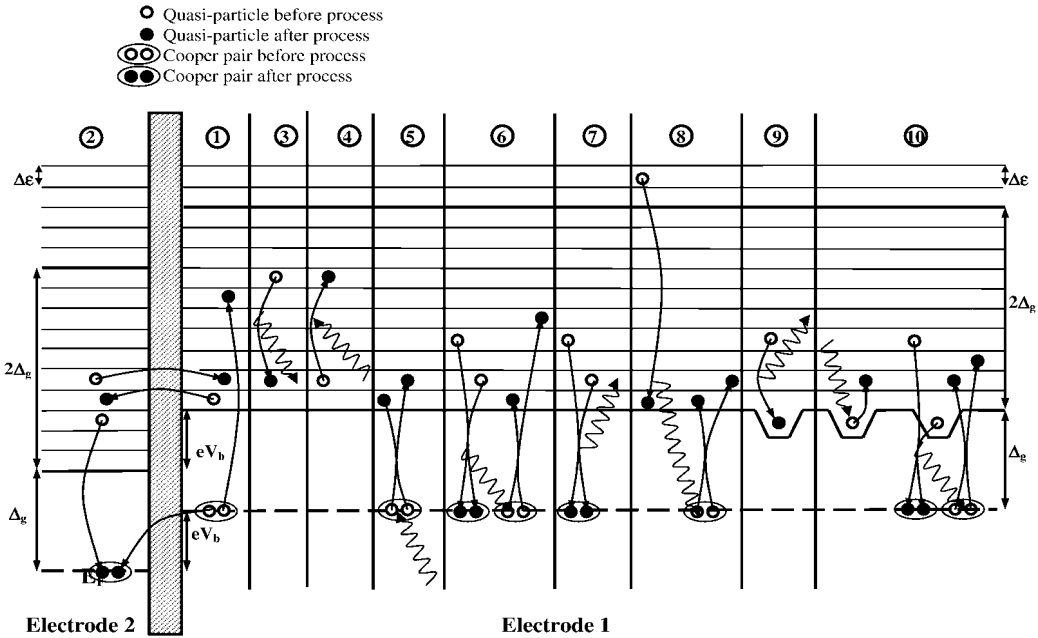

**Figure 1**. Schematical semiconductor representation of all processes included in our kinetic equations model. **1:** Tunnelling and back-tunnelling **2:** Cancellation tunnelling. The cancellation back-tunnelling is not shown for simplicity. It can be found by reversing the arrows on the back-tunnelling schematic of 1. **3:** Electron-phonon scattering with emission of a phonon (qp relaxation). **4:** Electron-phonon scattering with absorption of a phonon (qp excitation). **5:** Cooper pair breaking. **6:** qp recombination with energy exchange. **7:** qp recombination with phonon loss. **8:** qp multiplication. **9:** qp trapping by relaxation. **10:** qp de-trapping by phonon absorption and by recombination with an untrapped qp.

The rate equation for the qp population in the energy interval $\Delta\varepsilon_\alpha$ centred on the energy $\varepsilon_\alpha$ and in electrode i (i = 1,2), $N_i(\varepsilon_\alpha)$, can be written as [2]:

$$\frac{\partial N_i(\varepsilon_\alpha)}{\partial t} = I_{i,tun} + I_{i,qp-ph} + I_{i,rec} + I_{i,mult} + I_{i,trap} + I_{i,loss}, \quad (1)$$

where $I_{i,tun}$, $I_{qp-ph}$, $I_{i,rec}$, $I_{i,mult}$, $I_{i,trap}$ and $I_{i,loss}$ describe the rate of qp injection and extraction from the energy interval $\Delta\varepsilon_\alpha$ in electrode i due to qp tunnelling, qp-phonon scattering, qp recombination, qp multiplication, qp trapping and qp loss respectively. There is an equation similar to (1) for every single one of the $n_{en}$ energy intervals and for each of the two electrodes of the junction. These $2n_{en}$ equations have to be completed by the rate equations for the number of qps confined in the localised qp traps in electrode i, $N_{i,trap}$ [5]:

$$\frac{\partial N_{i,trap}}{\partial t} = -\sum I_{i,trap}, \quad (2)$$

which leaves us with a system of $2n_{en}+2$ coupled, non-linear differential equations to solve numerically. Except from the qp multiplication term, expressions for the contributions in Eqs. (1) and (2) can be found in Ref. [2] as a function of the energy dependent qp population $N_i(\varepsilon)$ in both electrodes of the detector. The model has five independent fitting parameters, which all have a profound physical meaning. These fitting parameters are the qp loss rate, the phonon escape rate from the electrode, the trapping probability, the number of available traps in the electrode and the trap depth. All these parameters can be determined independently by comparing simulations with the experimental photon energy- and temperature-dependent responsivity of the detector.

In this paper we will take a closer look at the qp multiplication term $I_{mult}$ and its effect on the response of the detector to photon absorption experiments. The qp multiplication term is described by the following sequence of events: A qp within the energy interval $\Delta\varepsilon_\alpha$ that has an energy larger than $3\Delta_g$ relaxes down to an energy interval $\Delta\varepsilon_\beta$ close to the gap energy $\Delta_g$ of the superconducting electrode. In this process it releases a phonon of energy larger than $2\Delta_g$, which in turn is able to break a Cooper pair and create two new qps in the energy intervals $\Delta\varepsilon_\gamma$ and $\Delta\varepsilon_\delta$ respectively (See also process 8 in Fig. 1). Because of the energy conservation law the energy interval $\Delta\varepsilon_\delta$ is fixed to being equal to the interval $\Delta\varepsilon_{\alpha-\beta-\gamma}$ centred on the energy $\varepsilon_\alpha$-$\varepsilon_\beta$-$\varepsilon_\gamma$. For electrodes consisting of a pure BCS superconductor this process rate per quasiparticle in the energy interval $\Delta\varepsilon_\alpha$, $\Gamma_{mult}(\varepsilon_\alpha, \varepsilon_\beta, \varepsilon_\gamma)$, is given by [6]:

$$\Gamma_{mult} = \frac{8\pi N_0 \alpha^2}{\hbar N} \frac{\delta\varepsilon^2}{\tau_0 (k_b T_C)^3} \frac{(\varepsilon_\alpha - \varepsilon_\beta)^2}{\Gamma_{PB}(\varepsilon_\alpha - \varepsilon_\beta)} \frac{\varepsilon_\alpha \varepsilon_\beta - \Delta_g^2}{\sqrt{\varepsilon_\alpha^2 - \Delta_g^2}\sqrt{\varepsilon_\beta^2 - \Delta_g^2}} \frac{(\varepsilon_\alpha - \varepsilon_\beta - \varepsilon_\gamma)\varepsilon_\gamma - \Delta_g^2}{\sqrt{(\varepsilon_\alpha - \varepsilon_\beta - \varepsilon_\gamma)^2 - \Delta_g^2}\sqrt{\varepsilon_\gamma^2 - \Delta_g^2}}$$
(3)

where $\Gamma_{PB}(\Omega)$ is the pair breaking time of a phonon of energy $\Omega$, $N_0$ is the single spin normal state density of states at the Fermi energy, N is the ion density and $\tau_0$ is the electron-phonon scattering characteristic time of the material defined in Ref. [4].

The multiplication term only has an effect when some of the qps in the electrodes reach the "active" region, which is the energy range above $3\Delta_g$. Qps only reach these high energy levels if qp relaxation is very slow, qp tunnelling is fast and the qp loss is slow. This effect can therefore only be observed in high quality, low energy gap junctions with very transparent tunnel barriers.

## 3. Development of high quality Al superconducting tunnel junctions

We have developed high quality Al STJs for use as photon detectors in astronomy [3]. All the junctions were fabricated by Cambridge MicroFab Ltd (Cambridge, UK). The 100 nm thick base electrode is DC-sputtered onto a sapphire substrate at liquid nitrogen temperature. Without breaking vacuum a thin (~1nm) Al oxide layer is grown on top of the film followed by the deposition of the 50 nm top electrode. The tri-layer is patterned using UV-lithography and covered by 350 nm of reactively sputtered Si oxide. The contacts to the top film and the 3x3 $\mu m^2$ plug in the base lead are made out of Nb in order to prevent diffusion of the qps out of the junction area. Fig. 2 shows a schematic cross-section of the junctions. The devices have a square geometry with lateral device sizes equal to 30, 50 and 70 $\mu$m.

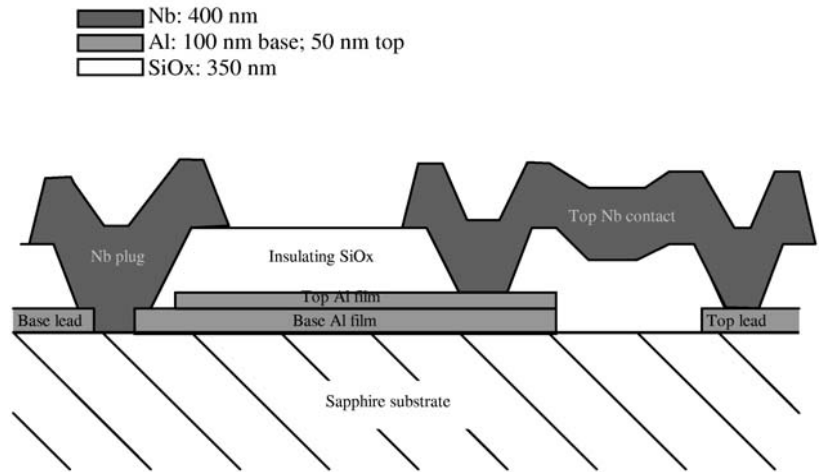

**Figure 2.** Schematic cross-section of an Al-based STJ showing from left to right: the Al base lead, the Nb base lead plug, the junction tri-layer, the Nb top contact and the Al top lead. The drawing is not to scale in the horizontal direction.

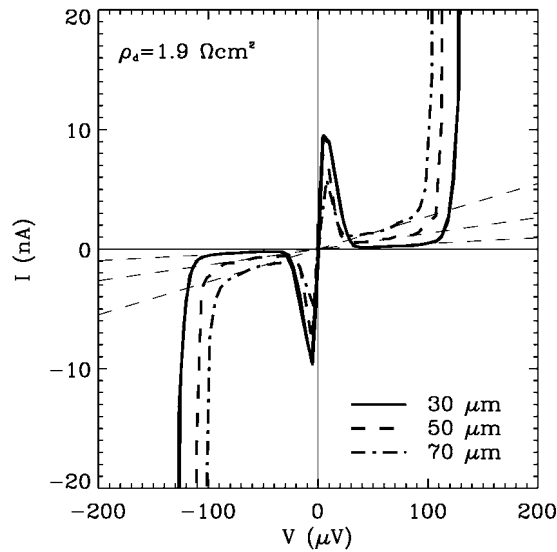

**Figure 3.** IV-curves of the sub-gap region of three junctions with side-lengths of 30, 50 and 70 $\mu$m. The thin dashed lines are fits to the dynamical resistances in the sub-gap regime.

The junctions are operated in a two-stage adiabatic demagnetisation refrigerator with a base temperature of about 35 mK. The junctions are coupled via an optical fibre to a Xenon lamp in combination with a double grating monochromator, able to produce photons in the wavelength range from 250 to 1000 nm. 6 keV photons from a radioactive $^{55}$Fe source installed 3-5 mm away from the junctions can be detected as well.

The responsivity and decay time of the pulses created by the photon absorption experiments are measured with the room temperature read-out electronics. A charge sensitive preamplifier integrates the current pulse created by the photo-absorption process in the junction. The output signal of the preamplifier is then fed into the shaping stage, consisting of two bipolar semi-gaussian shaping filters. In parallel to the photo-signal, an electronic pulser can be fed into the preamplifier, in order to measure the electronic noise.

The insulating barrier between the electrodes is uniform and of high quality. Fig. 3 shows an IV-curve of the sub-gap region of the 30, 50 and 70 μm side length junctions taken at 35 mK with an applied parallel magnetic field of 50 Gauss. Leakage currents are as low as 100 fA per μm$^2$ of junction area and the dynamical resistivity $\rho_d$ in the bias domain is equal to ~2 Ω cm$^2$. On the other hand the normal resistivity $\rho_{nn}$ of the barrier is equal to ~7 μΩ cm$^2$, giving a quality factor $Q=\rho_d/\rho_{nn}\sim 3\ 10^5$.

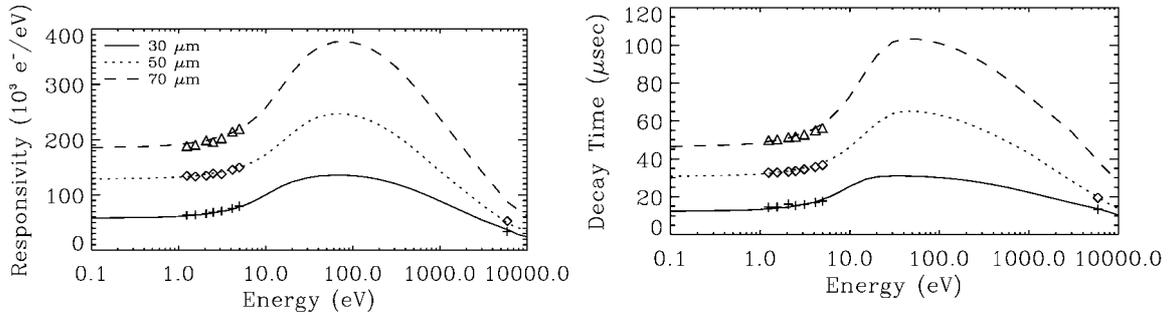

**Figure 4.** Measured responsivity and decay time of the 30 (crosses), 50 (diamonds) and 70 (triangles) μm Al junctions as a function of incoming photon energy for an applied bias voltage of 50 μV. Optical data is shown as well as 6 keV x-ray results. The results of the simulations are shown as well.

## 4. Experimental results and simulations made with the model

All the junctions were illuminated with NIR to soft UV photons (λ=250-1000nm). For every photon energy the responsivity and decay time of the current pulse were measured. Fig. 4 shows the experimental responsivity and decay time of the different Al junctions as a function of the incoming photon energy. In the figure the optical data is shown as well as the data for 6 keV soft x-ray detection experiments performed with the 30 and 50 μm side length junctions.

By varying the five fitting parameters of the model a fit to the experimental data in Fig. 4 was made for the three junctions. The results of the model are shown in the figure as well. The parameters which give the best fit to the data are shown in Table 1. The low number of qp traps in the junctions is striking: only ~7000 localised trapping states are available in the electrodes of the junctions, which is one and two orders of magnitude lower than in comparable Ta and Nb based junctions respectively.

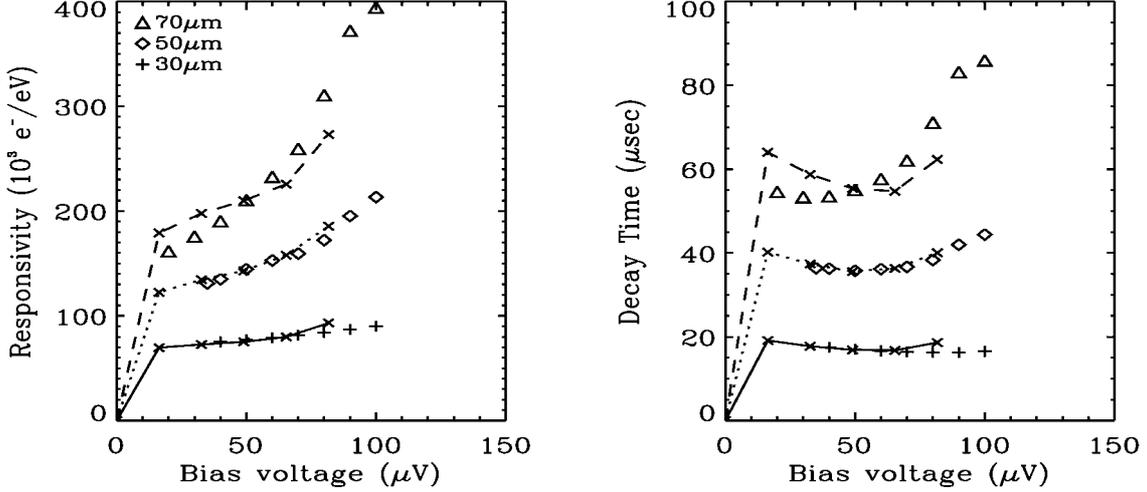

**Figure 5.** Responsivity and decay time as a function of applied bias voltage for the absorption of 300 nm (4.13 eV) photons in Al junctions. The experimental data is shown for the 30 (crosses), 50 (diamonds) and the 70 µm (triangles) side length junctions. The simulated data is shown as well for the 30 (solid line), 50 (dotted line) and the 70 µm (dashed line) devices.

**Table 1:** Fitting parameters of the model for the three Al junctions with side-lengths of respectively 30, 50 and 70 µm.

| Fitting parameter | Unit | 30 µm | 50 µm | 70 µm |
| --- | --- | --- | --- | --- |
| Quasiparticle loss time | µsec | 32 | 69 | 108 |
| Phonon escape rate | Hz | $1.4\,10^9$ | $1.4\,10^9$ | $1.4\,10^9$ |
| Trapping probability | / | 0.055 | 0.02 | 0.014 |
| Number of traps in electrode | / | 5300 | 7800 | 7700 |
| Trap depth | µeV | 81 | 81 | 81 |

The fact that the number of traps does not vary with the device size and that the trapping probability decreases with increasing device size indicates that the trapping centres are very localised and not in the bulk or at the perimeter of the Al junctions. The most probable position of the trapping centres is at the positions of the Nb contacts to the top and base electrodes.

The variation of the responsivity and the decay time with bias energy was measured as well for the absorption of 4.13 eV photons ($\lambda = 300$ nm). The experimental data is shown in Fig. 5 as well as the simulations performed with the kinetic equation model including the qp multiplication term and using the parameters of table 1. Note that simulations are only possible for bias energies which are an integer multiple of the basic energy interval $\delta\varepsilon$, because otherwise qps have to be split up in between intervals after tunneling, which introduces numerical errors. This explains the limited number of data points in the simulated curves. The responsivity and decay time show a strong increase with bias voltage, the dependence being strongest for the larger junctions with the longer loss times. In addition the increase of the responsivity is relatively stronger than the increase of the decay time with bias voltage. These very unusual dependencies are related to the qp energy distribution in the low energy gap, low loss junctions with fast tunnelling.

The simulations of the energy dependence of the qp population show that it converges within a microsecond to a "quasi-equilibrium" distribution. This qp energy distribution is called to be in "quasi-equilibrium" in the sense that the normalised distribution stays constant, whereas the absolute number of qps diminishes in the electrodes because of the different qp loss channels. Fig. 6 shows the quasi-equilibrium distribution 15 μsec after the absorption of a 500 nm photon in the 70 μm junction with four different bias voltages applied. The distribution gets broader with increasing bias voltage. For bias voltages larger than ~60 μV a certain fraction of the qps reach the region above $3\Delta_g$ and start the multiplication process. The qps created by this multiplication process slow down the loss process, which in turn increases the decay time of the pulse. This explains the unusual dependence of the decay time on applied bias voltage. If the bias voltage increases even more, the qp generation by the multiplication process outnumbers the qp losses. In this case the qp population diverges and photon detection experiments cannot be performed anymore. The bias voltage for which this occurs (~100 μV) corresponds to the bias voltage of the current step, which can be seen in the IV-curves of the devices (Fig. 2(b)). The exact bias voltage value for which the current step occurs depends on the loss time of the junctions and is therefore slightly different for the three device sizes. More details about IV-characteristics of high-quality lower gap junctions can be found in Refs. [1] and [6].

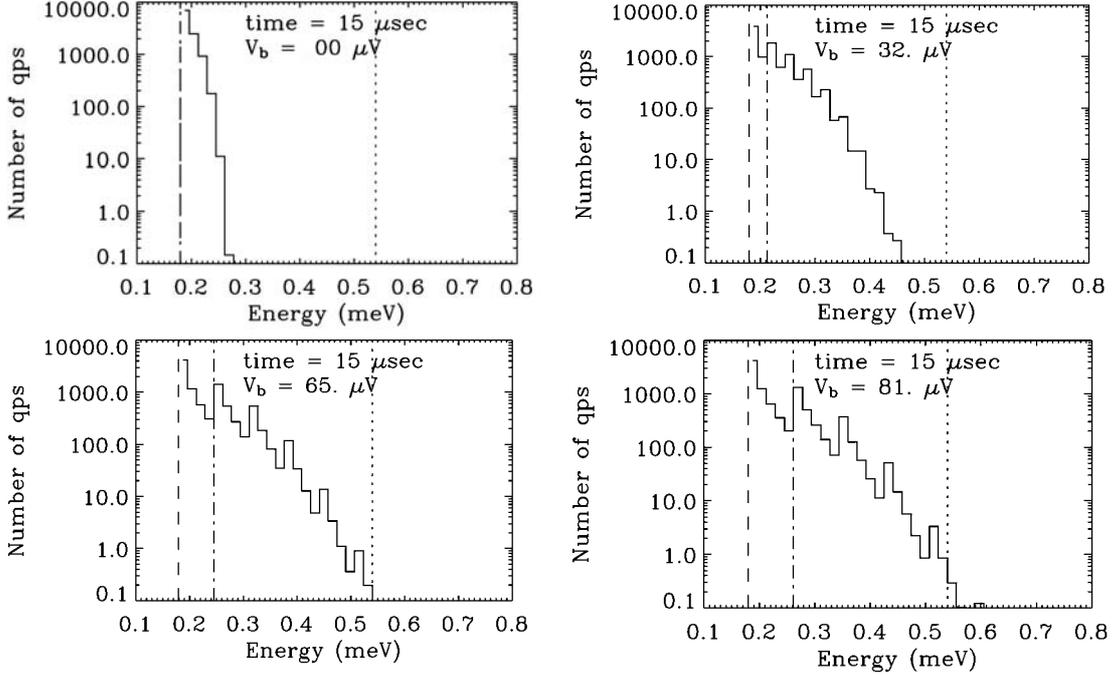

**Figure 6.** Variation of the quasiparticle energy distribution 15 μsec after the absorption of a 500 nm photon in a 70 μm side length Al junction as a function of the applied bias voltage. The quasi-equilibrium distributions of four different bias voltage settings are shown, which are respectively 0, 32, 65 and 81 μV. The vertical dashed line indicates the gap energy, the vertical dashed-dotted line indicates the bias energy above the gap and the vertical dotted line indicates the $3\Delta_g$ energy level, which is defined as the minimum energy of the active region.

The fact that the responsivity increases relatively faster with bias voltage than the decay time arises from a smaller fraction of charge lost via cancellation tunnel events. When the bias voltage is large, the fraction of qps which can undergo cancellation tunnel events is smaller than for smaller bias voltages. Note that this fraction is rather large in the lower gap junctions, because of the very broad energy distribution of the qps. Simulations show that the ratio of average number of tunnel events per qp to the average number of cancellation tunnel events per qp varies almost linearly between 1 at zero bias to 1.75 at 100 μV. This shows that even for an applied bias voltage of 100 μV, 60% of the created charge output due to tunnel events is annihilated by the cancellation tunnel events.

## 5. Conclusions

We have introduced the qp multiplication process in the framework of a time- and energy-dependent kinetic model of photon absorption by superconducting tunnel junctions. The model is applied to high quality Al junctions, which were tested as optical to x-ray photon detectors in a 35 mK environment. Simulations of the photon energy dependence of the responsivity and decay time of the detectors show that the number of localised qp trapping states is very low and independent of the device size of the detectors. It is concluded that most of the traps are therefore in the Nb of the contacts to the top and base electrodes. Both the responsivity and the decay time show a strong increase as the bias voltage increases. This unusual dependency can be explained by taking the very broad qp energy distribution into account, which forms due to successive qp tunnel events and incomplete energy down-conversion of the qps. The larger the bias voltage, the larger is the qp distribution. When the energy distribution is broad enough for the first qps to reach an energy level equal to $3\Delta_g$, qp multiplication sets in, which increases the decay time of the photo-pulse. The simulations performed with the energy dependent kinetic equations model agree very well with the experimental data.